\documentclass{PoS}
\usepackage{amsmath}
\usepackage{graphicx}
\pdfoutput=1

\title{Kaon semileptonic decay form factors from $N_f = 2$ non-perturbatively $O(a)$-improved Wilson fermions}  

\ShortTitle{Kaon semileptonic decay form factors}

\author{
QCDSF collaboration: D.~Br\"ommel$^{ab}$, R.~Horsley$^c$,
\speaker{S.M.~Morozov}$^d$, Y.~Nakamura,$^e$
D.~Pleiter$^e$, G.~Schierholz$^{ae}$, H.~St\"{u}ben$^f$, J.~Zanotti$^c$\\
\llap{$^a$}
Deutsches Elektronen-Synchrotron DESY, 22603 Hamburg, Germany\\
\llap{$^b$}
Institut f\"{u}r Theoretische Physik, Universit\"at Regensburg,
93040 Regensburg, Germany\\
\llap{$^c$} School of Physics, University of Edinburgh,
Edinburgh EH9 3JZ, UK\\
\llap{$^d$}
Institute for Theoretical and Experimental Physics,
Moscow, 117218, Russia\\
\llap{$^e$}John von Neumann Institute NIC/DESY Zeuthen,
D-15738 Zeuthen, Germany\\
\llap{$^f$} Konrad-Zuse-Zentrum f\"ur Informationstechnik Berlin,
14195 Berlin, Germany\\
E-mail:
\email{smoroz@itep.ru}
}

\abstract{
We present first results from the QCDSF collaboration for the kaon semileptonic
decay form factors at zero momentum transfer, using two flavours of
non-perturbatively $O(a)$-improved Wilson quarks. A lattice determination of
these form factors is of particular interest to improve the accuracy on the
CKM matrix element $|V_{us}|$. Calculations are performed on lattices with
lattice spacing of about 0.08 fm with different values of light and strange
quark masses, which allows us to extrapolate to chiral limit. Employing double
ratio techniques, we are able to get small statistical errors.}

\FullConference{The XXV International Symposium on Lattice Field Theory\\
July 30 - August 4 2007\\
Regensburg, Germany}

\begin{document}

\section{Introduction}

Recently high emphasis is placed on testing the unitarity of the
Cabibbo-Kobayashi-Maskawa (CKM) quark mixing matrix.
In particular, the unitarity  of the CKM matrix implies the
unitarity constraint of the first row
\begin{equation}
|V_{ud}|^2 + |V_{us}|^2 + |V_{ub}|^2 = 1 - \delta\,.
\label{eq:unitarity_constraint}
\end{equation}
According to the PDG~\cite{Yao:2006px} $\delta$ is equal to $0.0008(11)$.
The uncertainty is still quite substantial and has to be decreased.
As $|V_{ub}|$ is much less than unity, about half of the error of
$\delta$ comes from the uncertainty of $|V_{us}|$.
Since the kaon semileptonic decay rate is proportional to
$|V_{us}|^2 |f_+(0)|^2$, this matrix element can be determined by
combining experimental results for this decay rate and theoretical
calculations of the vector form factor at zero momentum transfer $f_+(0)$.

The kaon semileptonic decay form factors $f_+(q^2)$ and $f_-(q^2)$ are defined as
\begin{equation}
\langle \pi(p\,') | V_\mu | K(p) \rangle =
f_+(q^2)(p+p\,')_\mu + f_-(q^2)(p-p\,')_\mu\,,
\label{eq:general_form_factors}
\end{equation}
where $q^2 = (p - p\,')^2$ is the momentum transfer and
$V_\mu = \bar{s}\gamma_\mu u $ is the weak current.
It is also convenient to introduce the scalar form factor defined as
\begin{equation}
f_0(q^2) = f_+(q^2)\left(1+ \frac{q^2}{M^2_{K} - M^2_{\pi}}\xi(q^2)\right),\,\,
\text{and}\,\, 
\xi(q^2) = \frac{f_{-}(q^2)}{f_{+}(q^2)}\,.
\label{eq:scalar_ff}
\end{equation}

The form factor $f_+(0)$ at zero momentum transfer was estimated by Leutwyler and
Roos~\cite{Leutwyler:1984je} in 1984. They obtained the value
$0.961(8)$, which is still used as a reference value to extract $|V_{us}|$
from the experimental data. However, to estimate higher order terms in the
chiral perturbation theory (ChPT) expansion, they used a model of the wave function of the 
pseudoscalar meson. Recent ChPT calculations~\cite{BijnensOthers}
favour a slightly larger value of $f_+(0) = 0.984(12)$.
Thus, it is desirable to calculate $f_+(0)$ non-perturbatively and
lattice calculation provides such an opportunity.

Recently a lot of lattice groups reported calculations of
$f_+(0)$~\cite{Becirevic:2004ya,Becirevic:2004bb,Okamoto:2004df,
Tsutsui:2005cj,Dawson:2006qc,Boyle:2007wg,Antonio:2007mh,
Zanotti:lat07}.
Their results coincide with each other and are in agreement
with that of Leutwyler and Roos. For a recent review
see~\cite{Juettner:lat07}.

In this paper we present first preliminary results for
these form factors from $N_f = 2$
flavours of light dynamical, non-perturbatively $O(a)$-improved
Wilson fermions and Wilson glue.

\section{Lattice setup}

For our calculation we
use a gauge ensemble at $\beta = 5.29$, which corresponds to a lattice spacing
$a = 0.075$ fm,\footnote{For translation into physical units the Sommer parameter
$r_0 = 0.467$ fm is used.}
and $\kappa_{\rm sea} = 0.13590$, which corresponds to a pion mass
$M_\pi = 0.591(2)\,\mbox{GeV}$.
The lattice volume is $24^3\times48$ with spacial extent equal to $1.9\,\mbox{fm}$.
The correlation functions have been calculated on about 800 gauge
configurations using various source locations to reduce statistical
noise.
We identify the sea quark mass with the light quark mass and
the valence quark mass with the strange quark mass using
3 values $\kappa_s = 0.13485, 0.13530$
and $0.13570$.
The corresponding kaon masses are
$M_K = 0.780(11), 0.704(13), 0.629(14)\,\mbox{GeV}$.

\section{Correlators}

On the lattice we calculate two and three point functions, which are defined by
\begin{eqnarray}
C_{P}(t, \vec{p}) &=&
\sum_{\vec{x}} \langle O_{P,snk}(\vec{x},t)O^\dagger_{P,src}(\vec{0},0) \rangle
e^{-i\vec{p}\vec{x}} \xrightarrow[t\rightarrow\infty]{}
\frac{Z^*_{P,src} Z_{P,snk}}{2 E_P(\vec{p})} e^{-E_P(\vec{p}) t},\\
C^{PQ}_\mu(t,t\,', \vec{p}, \vec{p}\,') &=& 
\sum_{\vec{x},\vec{x}\,'}\langle O_{Q,snk}(\vec{x}\,',t\,')V_\mu(\vec{x},t)
O_{P,src}^\dagger(\vec{0},0)\rangle e^{-i\vec{p}\,'(\vec{x}\,'-\vec{x})-i\vec{p}\,\vec{x}} \\
&\xrightarrow[t,(t\,'-t)\rightarrow\infty]{}&
\frac{Z^*_{P,src} Z_{Q,snk}}{4 E_P(\vec{p}) E_Q(\vec{p}\,') Z_V}
\langle Q(p\,')|V_\mu^{(R)}|P(p)\rangle  
 e^{-E_P(\vec{p})\, t - E_Q(\vec{p}\,')(t\,'-t)},
\label{eq:cordef}
\end{eqnarray}
where $P$ and $Q$ denote either the $K$ or $\pi$ meson. The energy of meson $P$
($Q$) is denoted by $E_P(\vec{p})$ ($E_Q(\vec{p})$). The renormalised vector current, including the renormalisation
factor $Z_V$, is denoted by $V_\mu^{(R)}$. The overlap with the physical meson states is
given by $Z_{P,snk}$ and $Z_{P,src}$.

% [dp]  In fact also t_src > 0 is used
In the following we will assume the
point meson sources to be inserted at time $t_{src} = 0$
(thus $t_{src}$ is omitted in Eq.~(\ref{eq:cordef})) and point 
sinks are inserted at $t\,' = T/2$. As $t\,'$ is fixed the
$t\,'$ dependence of all quantities is ignored in the following.
For this choice of $t\,'$ the three point functions (and therefore
also the double ratios) are symmetric
with respect to $T/2$.
We make use of this property and average over both time ranges
to increase the precision of calculations.

Note that smearing of meson operators is
not used. This is because it leads to a momentum dependence of
the overlap $Z_{P,snk}(p)$ and $Z_{P,src}(p)$. See, for example,
Appendix A in Ref.~\cite{Bowler:1996ws} and the discussion below.

\section{Scalar form factor at $q^2_{max}$}

\begin{figure}[!ht]
\hspace*{-6mm}
\includegraphics[width=8.8cm]{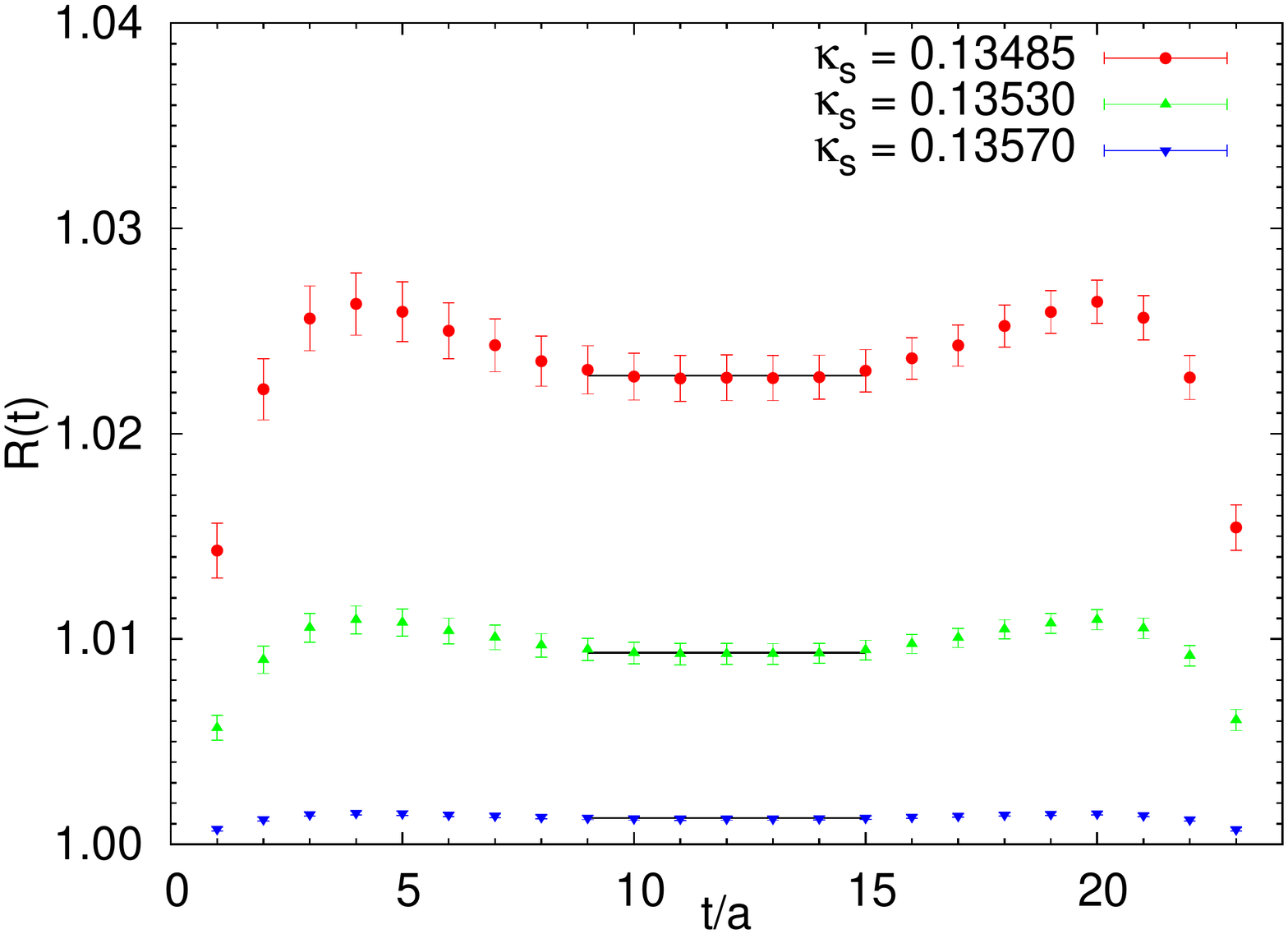} 
\hspace*{-12mm}
\includegraphics[width=8.8cm]{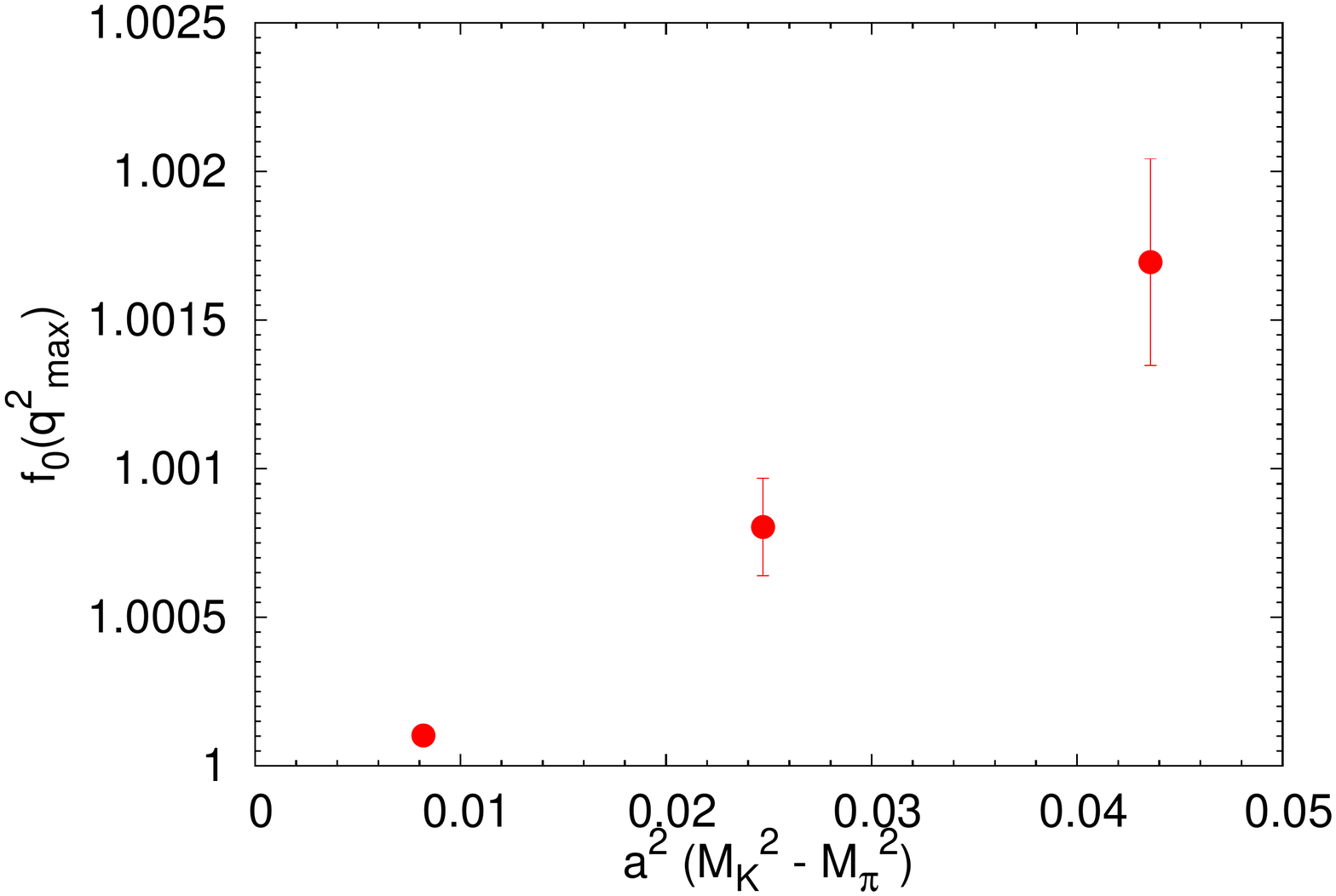}
\caption{{\bf Left:} Time dependence of the double ratio $R(t)$
(Eq.~({\protect\ref{eq:dr1}}))
for three values of the strange quark masses. {\bf Right:}
Values of {\protect $f_0(q^2_{max})$} as a function of the
$SU(3)$ breaking parameter $a^2\Delta M^2 = a^2(M_K^2 - M_\pi^2)$.}
\label{fig:dr1}
\end{figure}

The scalar form factor $f_0(q^2)$ at $q^2 = q^2_{max} = (M_K-M_\pi)^2$
can be obtained from the double ratio of three point functions (which
was originally proposed to calculate the $B\rightarrow Dl\nu$ form factor
in Ref.~\cite{Hashimoto:1999yp}):
\begin{equation}
R(t) = \frac{C^{K\pi}_4(t, \vec{0},\vec{0})
             C^{\pi K}_4(t, \vec{0},\vec{0})}{
             C^{K K}_4(t, \vec{0}, \vec{0})
             C^{\pi\pi}_4(t, \vec{0}, \vec{0})}
\xrightarrow[t\rightarrow \infty]{}
\frac{(M_K + M_\pi)^2}{4M_K M_\pi} (f_0(q^2_{max}))^2 = \bar{R}\,.
\label{eq:dr1}
\end{equation}
Our results for $R(t)$ for three values of the strange quark masses are shown 
in the left panel of Fig.~\ref{fig:dr1}. Note that renormalisation constants and exponential factors
exactly cancel in $R(t)$. We can extract $f_0(q^2_{max})$
with a statistical uncertainty $< 0.1\%$.

In the $SU(3)$ symmetric limit $\bar{R}$ is equal to unity. The deviation
of $\bar{R}$ from unity depends on the physical $SU(3)$ breaking effects on
$f_0(q^2_{max})$, which can be seen in the right panel of Fig.~\ref{fig:dr1}.

\section{Interpolation to zero momentum transfer}

To study the $q^2$ dependence of the form factor we calculate 
\begin{equation}
F(\vec{p},\vec{p}\,') = \frac{f_+(q^2)}{f_0(q^2_{max})}
\left(1 + 
\frac{E_{K}(\vec{p}) - E_{\pi}(\vec{p}\,')}
     {E_{K}(\vec{p}) + E_{\pi}(\vec{p}\,')}\xi(q^2)\right)
\end{equation}
from the second double ratio
\begin{equation}
R_F(t,\vec{p},\vec{p}\,') = \frac{
C^{K\pi}_4(t, \vec{p},\vec{p}\,') C^{K}(t,\vec{0}) C^{\pi}(T/2-t,\vec{0})}{
C^{K\pi}_4(t, \vec{0},\vec{0})        C^{K}(t,\vec{p}) C^{\pi}(T/2-t,\vec{p}\,')}
\xrightarrow[t\rightarrow\infty]{}
\frac{E_K(\vec{p}) + E_\pi(\vec{p}\,')}{M_K + M_\pi} 
F(\vec{p}, \vec{p}\,') = \bar{R}_F\,.
\label{eq:dr2}
\end{equation}

For meson point operators the overlaps cancel in (\ref{eq:dr2}). This is not true anymore 
if smeared meson operators are used, because their overlap with the physical state
depends on the momentum. That is why we are using point sources and sinks. A systematic study of this effect will be reported
elsewhere.

\begin{figure}[t]
\hspace*{-6mm}
\includegraphics[width=8.8cm]{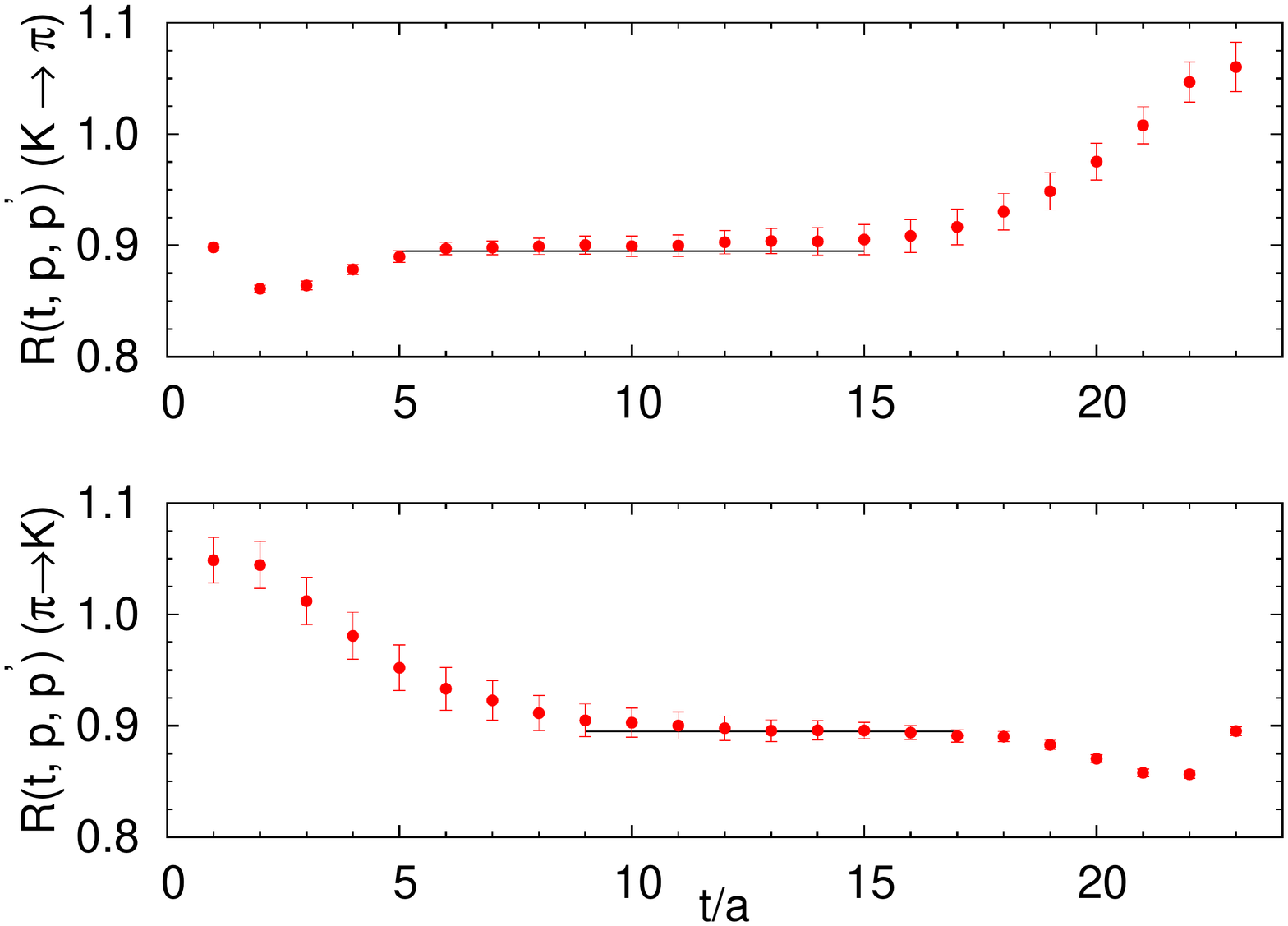}
\hspace*{-12mm}
\includegraphics[width=8.8cm]{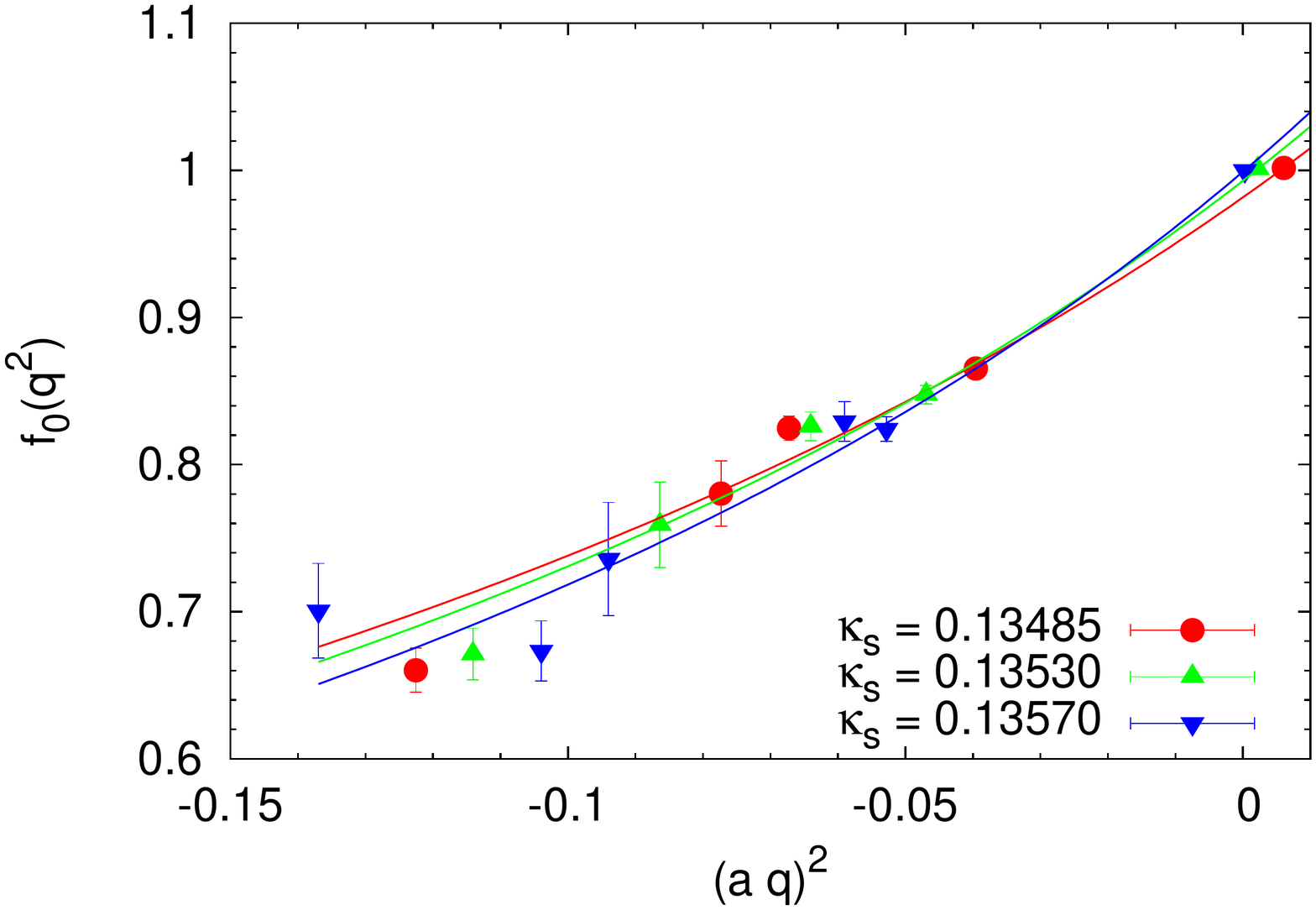}
\caption{{\bf Left:} Time dependence of double ratio {\protect $R_F(t,\vec{p},\vec{p}\,')$}
for $\kappa_s = 0.13485$. {\bf Right:}
Scalar form factor $f_0(q^2)$ for different values of the strange quark
mass. The solid line is the result of a monopole fit as described in the text.}
\label{fig:dr2_f0q2}
\end{figure}

The double ratio $\bar{R}_F(\vec{p},\vec{p}\,')$ is symmetric under exchange of $\pi$ and $K$
and their momenta, i.e.
$\bar{R}_F^{K\rightarrow\pi}(\vec{p},\vec{p}\,') =
\bar{R}_F^{\pi\rightarrow K}(\vec{p}\,',\vec{p})$,
which we make use of to increase the statistics at some values of $q^2$. In the left panel of
Fig.~\ref{fig:dr2_f0q2} we show, as an example, the data for
$R_F(t,\vec{p},\vec{p}\,')$ with $|\vec{p}\,'| = 0, |\vec{p}| = 1$ for $K\rightarrow \pi$ and
$|\vec{p}\,'| = 1, |p| = 0$ for the $\pi\rightarrow K$ transition.

In order to convert $F(\vec{p},\vec{p}\,')$ to $f_+(q^2)$, we need to
calculate $\xi(q^2)$. To do so, we define a third double ratio:
\begin{eqnarray}
R_k(t, \vec{p}, \vec{p}\,') = 
\frac
{C^{K\pi}_k(t, \vec{p}, \vec{p}\,')
 C^{KK}_4(t, \vec{p}, \vec{p}\,')}
{C^{K\pi}_4(t, \vec{p}, \vec{p}\,')
 C^{KK}_k(t, \vec{p}, \vec{p}\,')}\,\,\, ,\;\; k = 1,2,3\, .
\label{eq:dr3}
\end{eqnarray}
Then $\xi(q^2)$ is given by
\begin{equation}
\xi(q^2) = 
\frac
{-(\vec{p}+\vec{p}\,')_k (E_K(\vec{p}) + E_K(\vec{p}\,')) + 
  (\vec{p}+\vec{p}\,')_k (E_{K}(\vec{p}) + E_{\pi}(\vec{p}\,')) \bar{R}_k(\vec{p},\vec{p}\,')}
{ (\vec{p}-\vec{p}\,')_k (E_K(\vec{p}) + E_K(\vec{p}\,')) - 
  (\vec{p}+\vec{p}\,')_k (E_{K}(\vec{p}) - E_{\pi}(\vec{p}\,')) \bar{R}_k(\vec{p},\vec{p}\,')}\,,
\label{eq:xi}
\end{equation}
where $\bar{R}_k(\vec{p},\vec{p}\,') = \lim_{t\rightarrow\infty} R_k(t,\vec{p},\vec{p}\,')$.
The double ratio $\bar{R}_k(\vec{p},\vec{p}\,')$ is exactly unity in the $SU(3)$
symmetric limit and is sensitive to $SU(3)$ breaking effects
in the matrix element $C_\mu^{K\pi}$.

This ratio is the noisiest of all of them. Thus,
$\xi(q^2)$ could only be extracted with a large uncertainty. The relative error turns out to be 
$20\%-80\%$ for our statistics, even after we make use of the symmetry:
\begin{equation*}
\bar{R}^{K\rightarrow\pi}_k(\vec{p},\vec{p}\,') = 
\bar{R}^{\pi\rightarrow K}_k(\vec{p}\,',\vec{p}) = 
\bar{R}^{\pi\rightarrow K}_k(-\vec{p}\,',-\vec{p})\,.
\end{equation*}

From $f_0(q^2_{max})$, $F(\vec{p},\vec{p}\,')$ and $\xi(q^2)$ we can
calculate the scalar form factor $f_0(q^2)$. The results are shown
in the right panel of Fig.~\ref{fig:dr2_f0q2}.
To interpolate $f_0(q^2)$ to zero momentum transfer, we fit it with a
monopole ansatz $f_0(q^2) = f_0(0)/(1-q^2/M^2)$. The result of the fit is also shown in the figure.
 In the given range of momenta the data is well described by this ansatz.

\section{Chiral extrapolation}

In order to calculate the physical value of $f_+(0) = f_0(0)$ we have to
extrapolate our results to the
physical pion and kaon masses.
We make use of the results of ChPT
to guide our extrapolation.
In ChPT $f_+(0)$ can be expanded in terms of light pseudoscalar meson masses giving
\begin{equation}
f_+(0) = 1 + f_2 + f_4 + \dots\,\,\,\,,\,\, f_n = {\cal O}(M_{\pi,K,\eta}^{2n})\,.
\end{equation}
The leading correction $f_2$ receives only contributions from non-local operators and can be determined unambiguously
in terms of $M_K, M_\pi$ and $f_\pi$ (see~\cite{Becirevic:2005py}).
We compute $f_2$ at the actual pion and kaon masses and define
\begin{equation}
\Delta f = f_+(0) - (1 + f_2)\,,
\end{equation}
which receives only contributions from local operators.

\begin{figure}[t]
\hspace*{-6mm}
\includegraphics[width=8.8cm]{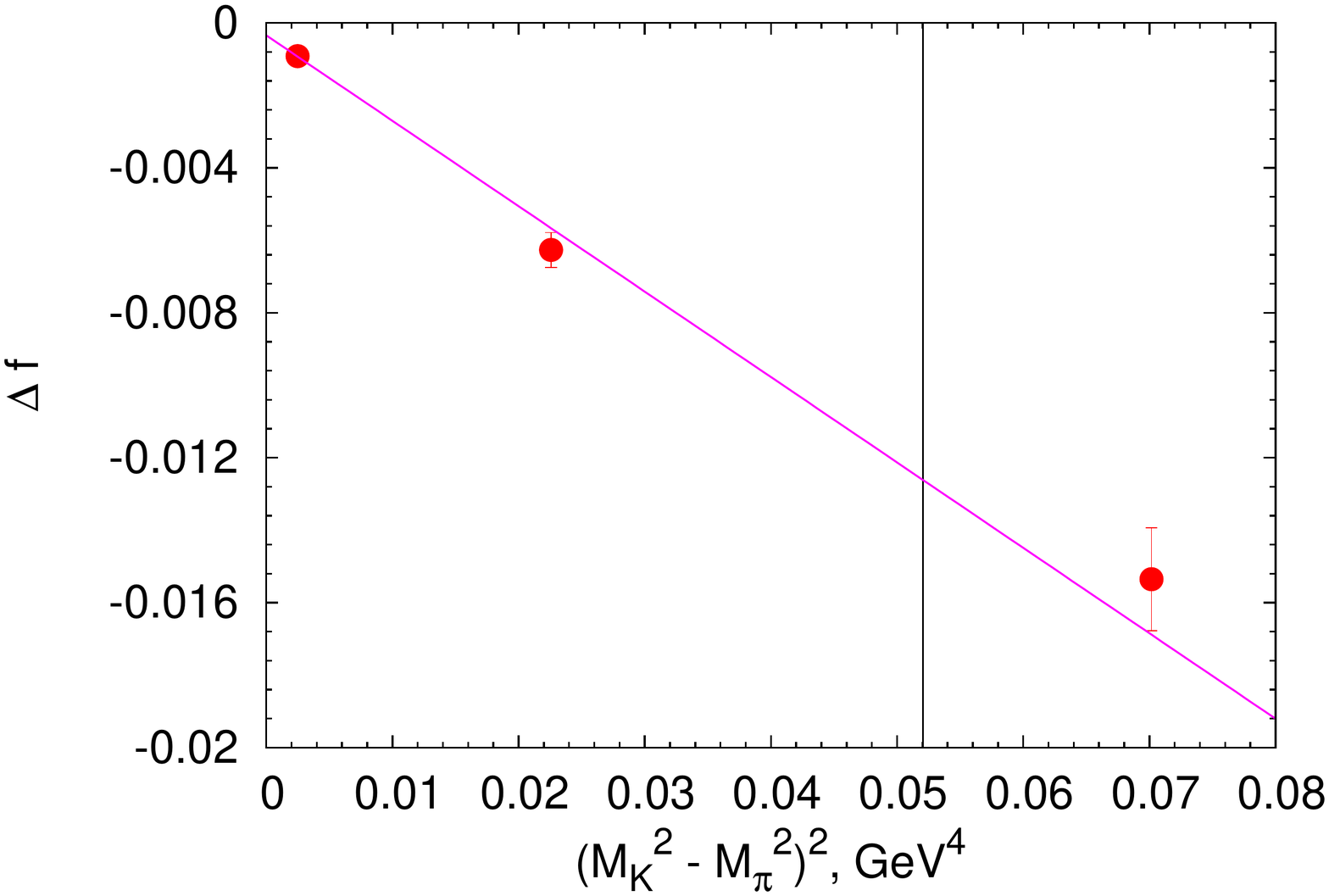}
\hspace*{-12mm}
\includegraphics[width=8.8cm]{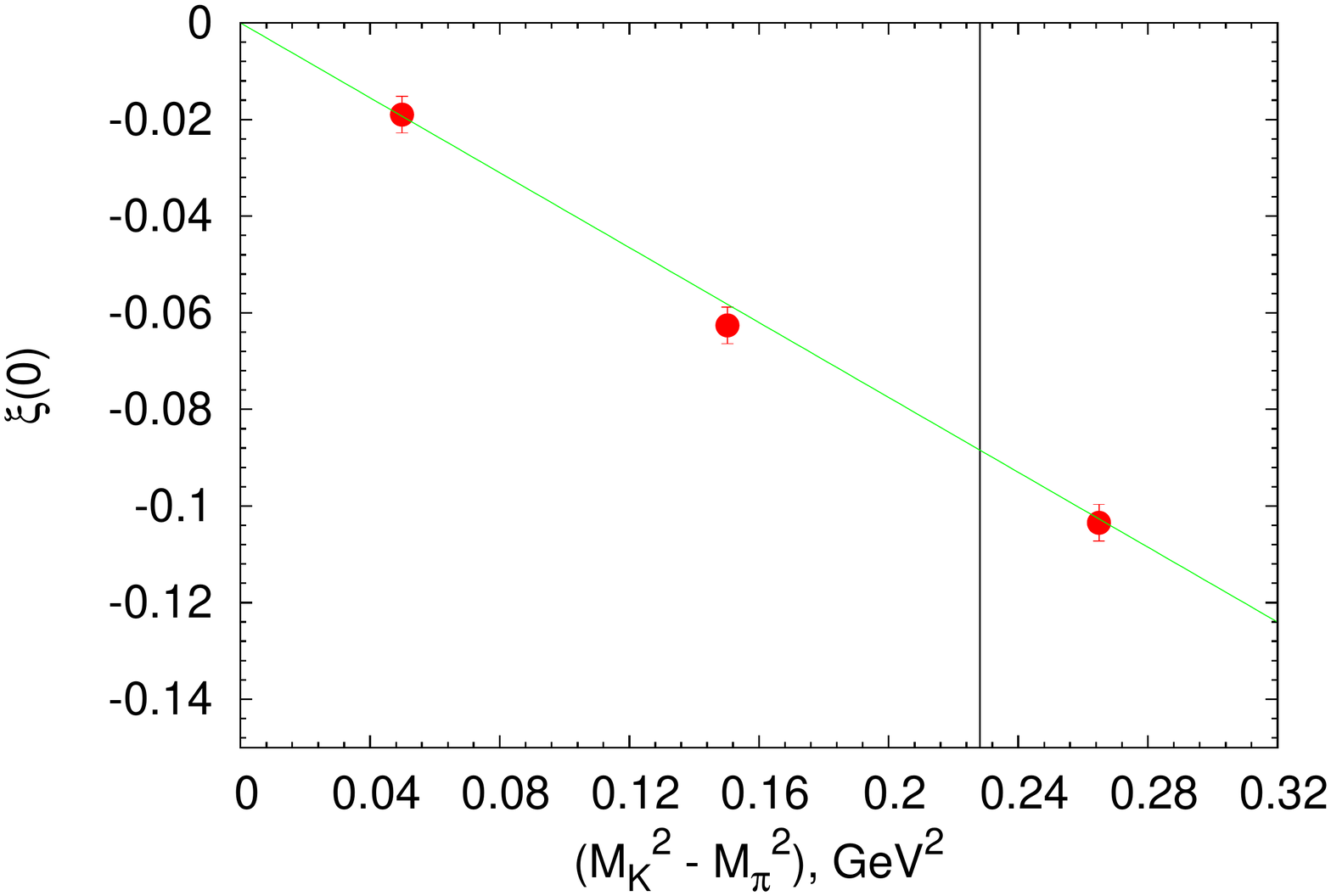}
\caption{
Interpolation of $\Delta f$ ({\protect {\bf Left}}) and $\xi(0)$
({\protect {\bf Right}}) to physical meson masses indicated by the vertical lines.}
\label{fig:chi_ext}
\end{figure}

The Ademollo-Gatto theorem~\cite{Ademollo:1964sr} states that $\Delta f$
is proportional to $(M_K^2 - M_\pi^2)^2$. Hence, we may write
\begin{equation}
\Delta f = a+b(M_K^2 - M_\pi^2)^2\,,
\label{AG}
\end{equation}
to obtain $f_+(0)$ by extrapolating (\ref{AG}) to the physical point. The
data points and the resulting fit are shown in the left panel of Fig.~\ref{fig:chi_ext}. 
We find $a \approx 0$, in agreement with the expected result $\Delta f = 0$ in the limit of flavour $SU(3)$.

At the physical meson masses we find $\Delta f = -0.0126(15)$~\footnote{Here and in the following only the statistical error is quoted.}.
This is to be compared
with $\Delta f = -0.016(8)$ of Ref.~\cite{Leutwyler:1984je}.
Inserting the physical value of $f_2$, $f_2 = -0.0227$, into $\Delta f$, we
obtain 
\begin{equation}
f_+(0) = 0.9647(15)_{stat}\,.
\end{equation}
We extract the CKM matrix element $|V_{us}|$ from the experimental value of
$|V_{us}|f_+(0)$ averaged over all decay modes~\cite{Moulson:2007fs},
$|V_{us}|f_+(0) = 0.21673(46)$. This finally gives
\begin{equation}
|V_{us}| = 0.2247(5)_{exp}(4)_{stat}\,.
\end{equation}

To compute $\xi(0)$ at the physical meson masses, we make the ansatz $\xi(0) =
c(M_K^2 - M_\pi^2)$, in accord with $\xi(0) =0$ in the $SU(3)$ limit. In the
right panel of Fig.~\ref{fig:chi_ext} we show the data points and the
resulting fit. We obtain
\begin{equation*}
\xi(0) = -0.10(2)_{stat}\,.
\end{equation*}
This value is consistent with the experimental values $-0.01(6)$ from
$K^0_{l3}$ decay and $-0.125(23)$ from $K^+_{l3}$ decay.

Note that in our analysis only statistical errors are estimated. Analysis of systematic
errors which include extrapolation errors of $f_0(q^2)$ to zero momentum transfer and
chiral extrapolation errors is underway.

\section{Conclusions and outlook}

%Note that the sub-percent level of precision of the experimental data
%requires lattice data of the same level of precision. We are confident that
%with our method of double ratios and recently proposed partially twisted
%boundary conditions~\cite{Boyle:2007wg} we will be able to meet this precision.
In this work we have presented preliminary results for the
kaon semileptonic decay form factors and $|V_{us}|$ from $N_f=2$ non-perturbatively
$O(a)$-improved Wilson fermions. We found 
\begin{equation}
f_+(0) = 0.9647(15)_{stat}\;,\;|V_{us}| = 0.2247(5)_{exp}(4)_{stat}\,, 
\end{equation}
in agreement with the results of other lattice
groups~\cite{Becirevic:2004ya,
Becirevic:2004bb,Okamoto:2004df,Tsutsui:2005cj,Dawson:2006qc,Boyle:2007wg,
Antonio:2007mh} as well as the estimate of Leutwyler and
Roos~\cite{Leutwyler:1984je}, within the error bars.

The QCDSF collaboration is going to improve the accuracy of the results
by performing calculations at lighter quark masses down to pion masses
of about $300$ MeV and using partially twisted boundary
conditions~\cite{Boyle:2007wg}. In addition, we are going to study
discretisation effects 
using gauge ensembles at different lattice spacings in the range of
$0.115$ fm to $0.070$ fm. Furthermore, we plan to study finite size effects.

\vspace{-2mm}
\section*{Acknowledgments}
\vspace*{-2mm}

The numerical calculations have been performed on the APE{\it 1000} and
apeNEXT at NIC/DESY (Zeuthen), the BlueGene/L at NIC/FZJ (J\"ulich) and
EPCC (Edinburgh). Some of the configurations have been generated on the
BlueGene/L at KEK by the Kanazawa group as part of the DIK research
programme.  This work was supported in part by the DFG, by the EU
Integrated Infrastructure Initiative Hadron Physics (I3HP) under contract
number RII3-CT-2004-506078. SMM is partially supported by the RFBR grants
05-02-16306a, 07-02-00237a, 06-02-04010 and INTAS YS Fellowship
05-109-4821. SMM also acknowledges support from the Lattice 2007 organizers.

\end{document}